\documentclass[aps,a4]{revtex4} 
\usepackage{graphicx} 
\usepackage{dcolumn}    
\usepackage{bm}             
\usepackage{gensymb}    
\usepackage{amsmath, amssymb, graphics}

\begin{document}

\title{Evidence for Dynamic Excitation-Inhibition Ratio in Networks of Cortical Neurons}
\author{Netta Haroush and Shimon Marom}
\affiliation{Technion --- Israel Institute of Technology, Haifa 32000, Israel. 
}

\begin{abstract}

\noindent In this report trial-to-trial variations in the synchronized responses of neural networks are offered as evidence for excitation-inhibition ratio being a dynamic variable over time scales of minutes. Synchronized network responses to stimuli were studied in \textit{ex-vivo} large scale cortical networks.  We show that sub-second measures of the individual synchronous response, namely -- its latency and decay duration, are related to minutes-scale network response dynamics.  Network responsiveness is reflected as residency in, or shifting amongst, areas of the latency-decay plane.  The different sensitivities of latency and decay durations to synaptic blockers imply that these two measures reflect the effective impacts of inhibitory and excitatory neuronal populations on response dynamics. Taken together, the data suggest that network response variations under repeated stimuli result from excitation-inhibition ratio being a dynamic variable rather than a parameter. 

\end{abstract}
\maketitle 

\section*{Introduction}
\noindent Evoked transient synchronous activity is acknowledged as significant in both normal and pathological neural conditions \cite{Uhlhaas2009}.  In mammalian brains, as well as in their \textit{ex-vivo} reduced experimental preparations (slices, cultured networks), an evoked transient synchronous activity has a temporally-stretched spike like shape in the population firing rate trace, a characteristic time scale (ca.~100 msec), the flavor of a threshold-governed event and a refractory period that lasts several seconds \cite{slovin2002,Derdikman2003,Eytan2006,Gullo2010,Weihberger2013}. With these properties in mind we adhere to a previously offered terminology and use the name Network Spike (NS) to designate evoked transient synchronous activity \cite{Eytan2006,Shew2009}. Trial-to-trial variations in the occurrence and the fine structure of stimulus evoked NSs were reported in both anesthetized and behaving animals as well as in reduced \textit{ex-vivo} preparations  \cite{Vogels1989,Snowden1992,Arieli1996,Weihberger2013,Shahaf2008}. The response variations reflect a multitude of factors that determine the network excitability status at the time of stimulus arrival. These include neuronal and synaptic noise, refractoriness of neuronal and synaptic activities and the context of ongoing activity within which a stimulus is applied \cite{faisal2008noise,Weihberger2013,Arieli1996,Fox2006}.

Here we took advantage of a relatively controlled experimental approach to large-scale cortical networks developing \textit{ex-vivo}, in order to expose the nature of response variations under repeated input over an extended range of time scales (from milliseconds to many minutes).  We provide indications for the existence of a hierarchy of timescales in the structure of trial-to-trial response variations, ranging from sub-seconds to many minutes.  We then show that response dynamics over minutes are reducible to the interplay of two instantaneous (i.e.~single NS, sub-second scale) observables: (1) the \textit{latency} from stimulus to the peak of the NS firing rate envelop, and (2) the \textit{decay} duration from that peak to baseline activity.  We show that these two instantaneous measures of the network excitability state---latency and decay duration---are differentially sensitive to specific pharmacological blockers of inhibitory and excitatory synaptic transmissions, suggesting that long-term network response variations reflect a dynamic excitation-inhibition ratio.  
 
\section*{Materials and Methods}

\subsection*{Cell preparation}
\noindent Cortical neurons were obtained from newborn rats (Sprague-Dawley) within 24 hours after birth using mechanical and enzymatic procedures described in earlier studies \cite{Marom2002}. The neurons were isolated and plated directly onto substrate-integrated multi electrode arrays. They were allowed to develop into functionally and structurally mature networks over a period of 2 weeks and were used in experiments within the period of 2--6 weeks post plating. The number of plated neurons was in the order of 450,000, covering an area of about $380$ mm$^2$ with heat-inactivated horse serum ($5\%$), glutamine ($0.5$ mM), glucose ($20$ mM), and gentamycin ($10~\mu$g/ml), and maintained in an atmosphere of $37^o$C, 5\% CO$_2$ and 95\% air in an incubator as well as during the recording phases.  An array of 60 Ti/Au extracellular electrodes, $30~\mu$m in diameter, spaced $500~\mu$m from each other (MultiChannelSystems, Reutlingen, Germany) was used. The insulation layer (silicon nitride) is pretreated with polyethyleneimine (Sigma, $0.01\%$ in $0.1$ M Borate buffer solution).

\subsection*{Electrophysiology}
\noindent A commercial amplifier (MEA-1060-inv-BC, MCS, Reutlingen, Germany) with frequency limits of 150--3,000 Hz and a gain of x1024 was used for obtaining data. Data was digitized using an acquisition board (PD2-MF-64-3M/12H, UEI, Walpole, MA, USA). Each channel was sampled at a frequency of 16 kHz, and detects electrical activity that might be originated from several sources (typically 2--3 neurons) as the recording electrodes were surrounded by several cell bodies. We have used a Simulink-based software for on-line control of data collection (see Zrenner et.al. \citeyear{Zrenner2010} for more details). Voltage stimulation was applied in the form of a mono-phasic square pulse $200~\mu$sec 800--950 mV through extracellular electrodes using a dedicated stimulus generator (MCS, Reutlingen, Germany).  Action potentials timestamps were detected on-line by threshold crossing of negative voltage. Detection of NSs was performed off-line using a previously described algorithm \cite{Eytan2006} based on threshold crossing of the network firing rate (binned to 3 msec).

\subsection*{Pharmacology} 
\noindent Inhibitory synaptic transmission was blocked with Bicuculline-Methiodide (Sigma-Aldrich) that was incrementally added to the bathing solution (final concentrations used: 0.5, 1, 1.5, 2, 2.5, 3, 4, 5 and 7 $\mu$M). 
Excitatory synaptic blocker (APV; amino-5-phosphonovaleric, Sigma-Aldrich) was added to networks that are already under a Bicuculine blockade.  Specifically, APV was added to networks that respond to stimuli in (or close to) a 1:1 manner, and where further application of Bicuculline did not change their response probability (5--12 $\mu$M Bicuculline). Final APV concentrations used were 50, 150 and 300 $\mu$M.  In all the experiments with pharmacological manipulations the stimulation rate was chosen to be 1/5 sec$^{-1}$ or slower, thus maintaining high response probabilities; once tuned, the stimulation rate was kept fixed for each network throughout the experiment.

\subsection*{Data Analysis }
\noindent \textit{Firing rate histograms and responsiveness}. Once a NS was detected within 1 second following a stimulus, action potentials recorded in all the electrodes within 1500 ms following the stimulus were extracted. Post-stimulus time histograms were constructed using a 1 ms time bin, and smoothed with a 5 ms moving average.  The \textit{latency} measure was defined as the time between stimulus onset and the first maximum of the smoothed firing rate histogram. To compute the \textit{decay} duration, the firing rate histogram was further smoothed by a wider window of 30--100 ms, thus avoiding impacts of oscillations within that phase on the measure. Decay duration was defined as the time between the histogram first maximum and the first drop below 0.15 spikes/ms along the falling phase of the NS.  

\textit{Fano Factor}. Network activity was represented as a point process (NS trains, composed of absolute detection times, i.e. Ð not referenced to stimulation times). The variance of the network activity under repeating stimulation over extended time periods (24 hours) was estimated by calculating its Fano factor \cite{Scharf1995,Lowen:1996pf}.  Count sequences Z(T(n)) were calculated using logarithmically spaced bin sizes T(n). For each count sequence the Fano factor, FF(T), defined as the variance of Z(T) divided by the mean of Z(T), was plotted as a function of the bin size T. The same analysis was applied to a surrogate data set generated by randomizing the inter-NS-intervals. The final rising section of the FF curves was fitted to a power law ($\alpha T^\beta$).

\textit{Autocorrelation}. The time series of latency and decay duration were used to calculate the autocorrelation of these processes, as if they were equally spaced in time, resulting in a warped time axis. Therefore, the autocorrelation was computed for ``serial  index lags" (see Figure {\bf 5A}), instead of actual time lags, according to 
$\sum _{i=1}^{n-l} \frac{(x_i-\hat{\mu }) (x_{l+i}-\hat{\mu })}{\sum _{i=1}^n (x_i-\hat{\mu }){}^2}$
Where the lag $l$ is the difference between the serial indexes, in the stream of successful responses.

\section*{Results}

\noindent The basic experimental procedure involved recording of activity while repetitively stimulating the network from a single site with constant amplitude at a constant rate. The data reported here originate from experiments ($n = 27$ networks) in which the stimulation amplitude was between $800$ to $950$ millivolts, delivered at a constant rate in the range of $1/3$ to $1/12$ sec$^{-1}$ (in different preparations and/or experimental sessions).  The data were reduced to a series of network spike time-stamps using a threshold-based criterion for identification of synchronous activity (evoked as well as spontaneously occurring; see methods).

The wide range of time scales present in neural population activity under stimulation is shown in Figure 1:  Panel A demonstrates a typical evoked network spike; the latency from stimulus to the peak of the network spike, as well as the decay of activity from that peak, are processes that take place within the tens and hundreds of milliseconds, constituting the shortest timescales in our neuronal population data. The next timescale in the hierarchy is the network spike refractory period, which is the minimal time between two consecutive network spikes, ranging between 1--10 seconds \cite{Robinson1993,Maeda1995a,Eytan2006}. The network refractory period is thus more than one order of magnitude longer compared to the above mentioned characteristic latency and decay times.
\begin{figure}[h!]
\begin{center}
\includegraphics[trim=0 0 0 0,width=15cm]{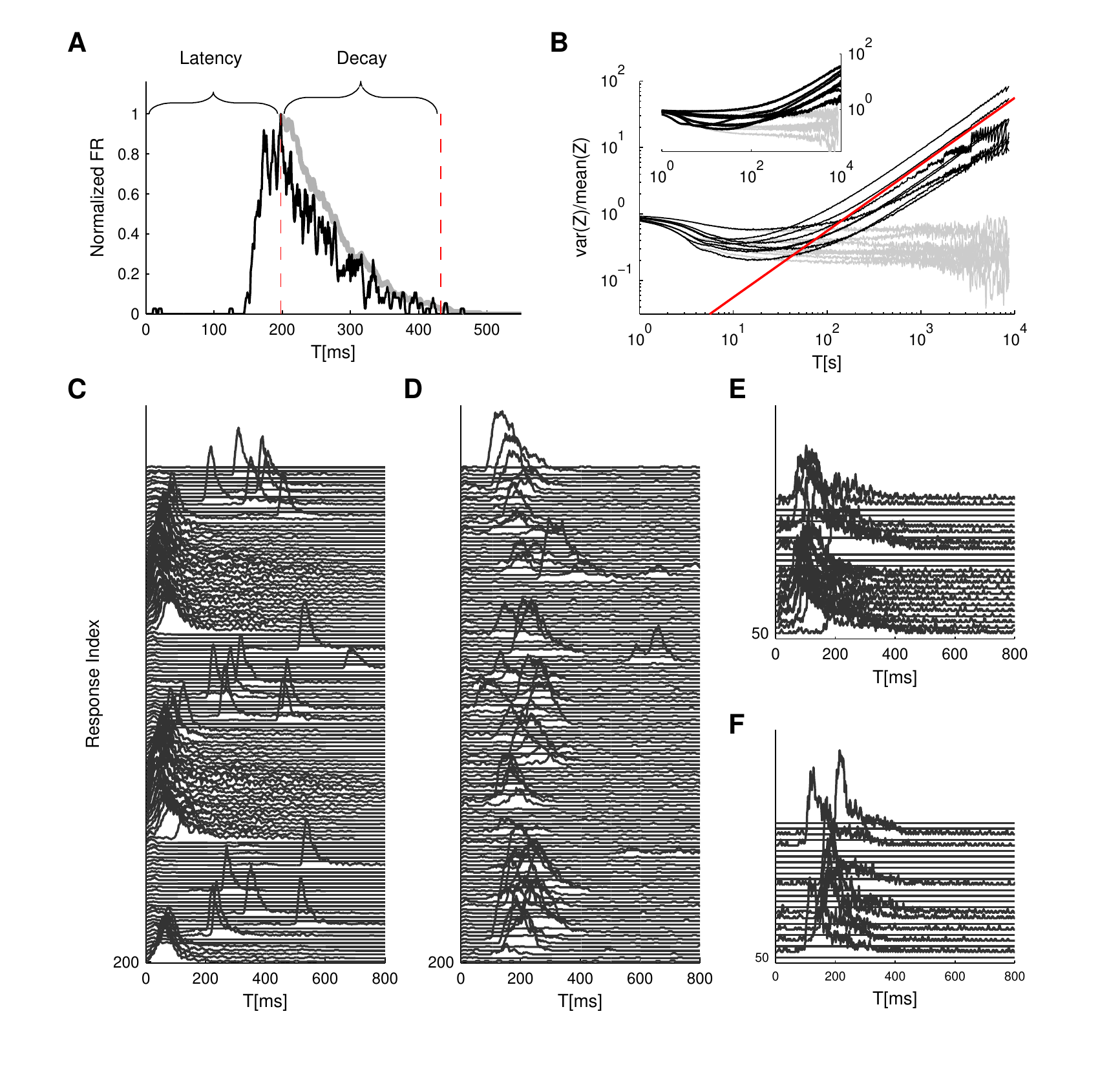}
\end{center}
\caption{\small{Time hierarchy of network activity: Panel {\bf A} shows a typical response to a single trial. The black trace depicts the network firing rate, binned at 1 ms and smoothed by a 5 ms moving average; the gray trace is a further smoothed version (30 ms moving average, in this case) of the falling phase of the NS, used to evaluate the \textit{decay duration}. Both traces are normalized to their maximal firing rate. Panel {\bf B} depicts the variance over mean (i.e. Fano factor) of NSs count ($Z$) as a function of the integration time bins ($T$); data were collected from 8 networks that were continuously stimulated over 24 hours at 1/5 sec$^{-1}$. The Gray lines depict identical analysis of the surrogate data set. The red line designates a power law with a unity exponent for reference purposes. The inset of panel {\bf B} shows the same analysis for the cases of spontaneously evoked NSs (seven networks, 24 hours time series). Panels {\bf C} and {\bf D} demonstrate consecutive network responses, displaying two examples from a spectrum of observed response modes (every third response was removed to enhance visual clarity). Panels {\bf E} and {\bf F} demonstrate two response modes observed at the same network within the first hour of stimulation, with {\bf E} resembling the short-latency segments of {\bf C}, and {\bf F} resembling {\bf D}; in this particular network, response mode sharply switched around the middle of the first recording hour (every other response is removed to enhance visual clarity).}}
\label{fg:ff24h}
\end{figure}

To identify further, slower timescales in the dynamics of network spikes under ongoing stimulation at a constant rate, we resorted to a measure of statistical variance that was introduced by Teich and colleagues for the analysis of spike number distributions \cite{Teich1992a}. Specifically, time series of network spike occurrence were sliced to bins of $T$ seconds durations, and the variance-to-mean ratio of event counts, a.k.a.~Fano factor, was calculated as a function of $T$.  Figure 1B shows the results of this analysis; the data originates from eight different networks (black lines) that were stimulated for 24 hours at a rate of $1/5$ sec$^{-1}$. The gray lines depict results of the same analytic procedure, applied to shuffled, surrogate data sets. At the very high temporal resolutions the Fano factor goes to unity, reflecting the Poisson nature of probability to capture an event within short time bins; this segment has no physiological significance. As $T$ approaches the timescale of one second, the Fano factor starts declining, as the uniquely defined refractory period introduces regularity into the time series. (It is acknowledged the the stimulation cycle time is also a determinant at this temporal scale, but see below described analysis of spontaneous activity.) Should refractoriness be the sole source of temporal complexity, one would expect that the Fano factor will approach zero as $T$ further increases, beyond the seconds scale.  Apparently this is not the case:  as the temporal resolution further decreases (i.e.,~larger $T$), the Fano factor steadily increases. The slopes at the right hand straight segments of the Fano factor curves (lower temporal resolutions) follow a power-law with an averaged exponent value of $0.64\pm0.14$, as if increasingly longer temporal structures are revealed by integrating the activity over longer temporal scales. The inset of Figure 1B demonstrates a similar phenomenon in the spontaneous activity recorded form 7 other networks over 24 hours; the average exponent value here was $0.89\pm0.0.21$.

The wide spectrum of these temporal structures may be intuited by observing the traces in Figure 1C--F. Panels C--D, recorded from two different networks, demonstrate very different responsiveness patterns: one having a rather cyclic nature (C) and the other is more irregular (D); in these examples the responsiveness patterns of networks were stable throughout  50 minutes of recordings. As implied by the analyses shown in Figure 1B, the response patterns exhibited by a given network change over time. The examples presented in panels E–-F were obtained from a network that switched its responsiveness pattern, during the first half of recording hour, from a mode resembling that of panel C (demonstrated in panel E) to response mode more similar to that presented in panel D (demonstrated in panel F).

\begin{figure}[h!]
\begin{center}
\includegraphics[trim=0 0 0 0,width=15cm]{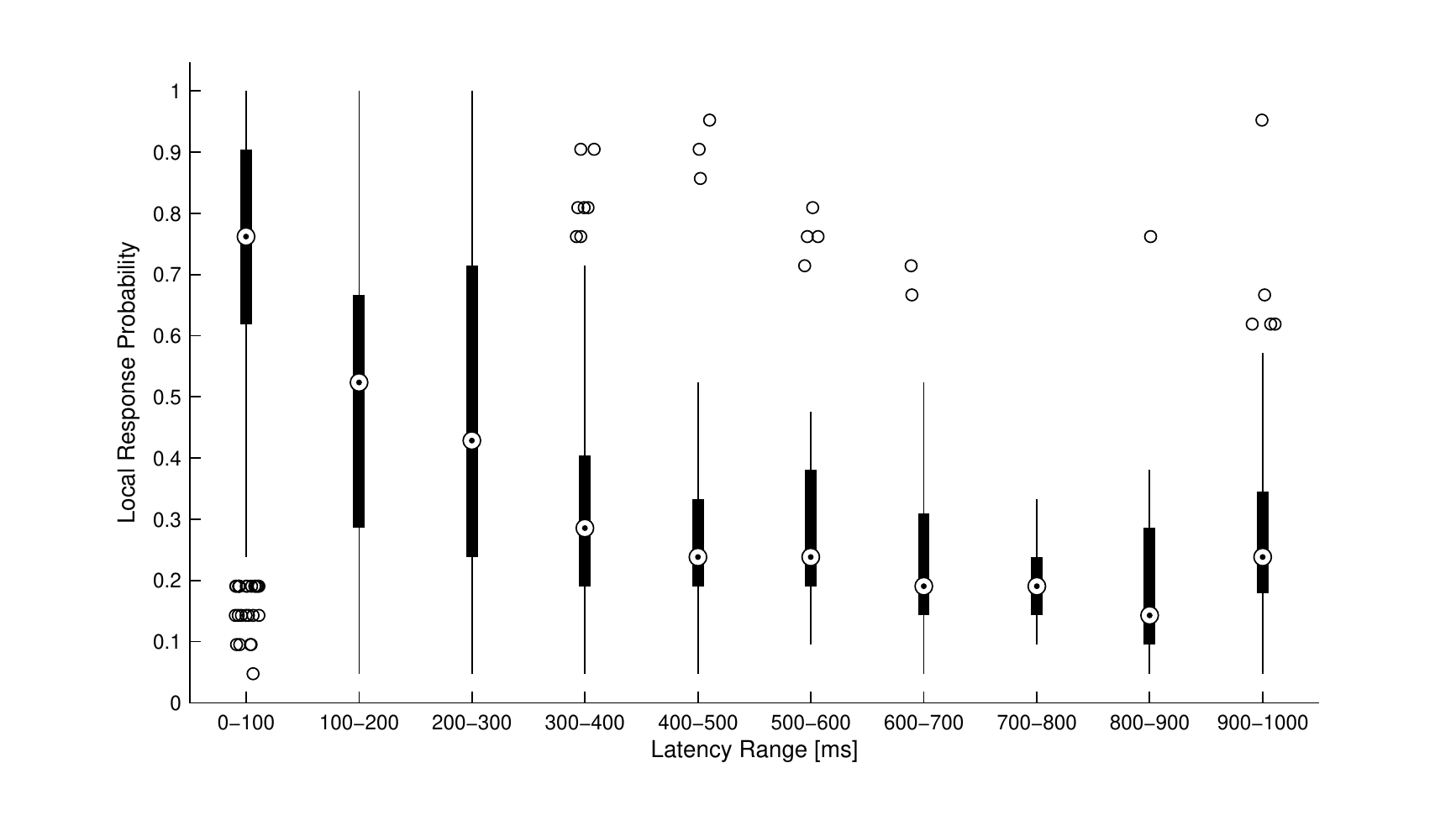}
\end{center}
\caption{\small{Local response probability correlates with response latency:
Local response probability distributions were grouped according to their latency range in a data set of 4146 responses pooled from 17 networks stimulated at 1/3--1/5 sec$^{-1}$ over 50 minutes. The upper and lower edges of each box mark the 75 and 25 percentiles of the data; the circle within the box represents the group's median. Whiskers are extended up to an equivalent of $\pm2.7$ standard deviations of each group, and individual outliers are marked by circles beyond the whisker range (49 outliers out of 4146 NSs). An inverse relation between the local response probability and the response latency emerges from the pooled data.}}
\label{fg:LatencyVsResP}
\end{figure}

The analyses described above suggest that the temporal structure of series of network spikes under repeated stimuli markedly deviates from an independent and identically distributed process. While the limit of refractoriness is reasonably understood and well documented in the literature \cite{Robinson1993,Maeda1995a,Weihberger2013}, the origin of correlations between evoked responses beyond the few seconds scale is poorly understood and, to the best of our knowledge, not described.  In what follows we focus on response dynamics within the range of 10--1000 seconds.

Many independent sources might contribute to the above dynamics of network responses over minutes. The data shown in Figure 1C--F suggest that these multiple sources may be lumped, at least coarsely, to processes that impact on the \textit{latency} from the stimulus time to the peak of the network spike: longer latencies seem to correlate with periods of lower response probabilities. To explore the relations between network spike latency and network responsiveness, we defined an instantaneous measure, the local response probability. Specifically, for each evoked network spike we calculated the fraction of responses to stimuli in the vicinity of that network spike through a symmetric window of 10 preceding and 10 following stimulation events.  Figure 2 shows the relations between local response probability and response latency; longer latencies are associated with lower local response probabilities (4146 evoked NSs collected from 17 networks, over an hour of repeated stimuli at 1/3--1/5 sec$^{-1}$). 
As previously reported studies show that features characterizing the envelope of the single spontaneously occurring (i.e., not stimulation evoked) network spike are sensitive to blockers of inhibitory synaptic transmission \cite{Eytan2006,Gullo2010}, we tested the relations between bath concentration of Bicuculline (an antagonist of GABA receptors) and network responses to stimuli. 
We gradually increased the drug's concentration from 0.5 $\mu$M up to 7 $\mu$M and in each concentration we exposed the network to a series of 40 stimuli delivered at a low constant rate (within the range of $1/5$--$1/12$~sec$^{-1}$).  Figure 3 summarizes the results of these experiments (six networks). Overall, application of Bicuculline resulted in increased network response probability (Figure 3D), and decreased latency (Figure 3E). In addition -- as may be seen in Figure 3F -- gradual block of inhibitory synapses results in increased decay duration. Networks displaying baseline activities marked by high response probability and short latencies were not significantly affected by Bicuculline application (data not shown). 

It is taken for granted that the coupling between network excitatory and inhibitory activities is such that, based on the results of Figure 3, no claim may be made regarding exclusive effects of Bicuculline on inhibition, as application of the drug may indirectly result in enhanced excitation. To expose -- even if partially -- the contribution of excitatory resources to network responsiveness, we applied increasing concentrations of APV, an NMDA antagonist, in the presence of high concentration (5 $\mu$M up to 12 $\mu$M) of Bicuculline. Under these conditions the responsiveness of the network is effectively determined by excitatory synaptic transmission. Figure 4 summarizes the effect of this experimental procedure (four networks).  Within the applied range of APV dosage (50--300 $\mu$M), response probability remained more or less stable and the response latency was practically unaltered; but the network spike decay duration became significantly shorter. As might be expected, considering the network spike as an excitable event of the network, higher doses of APV ($>$ 300 $\mu$M) induced a decrease of response probability accompanied by an increase of response latency (data not shown). The overall picture emerging from the experiments of Figures 3 and 4 is that response latency is more sensitive to the availability of inhibition, whereas decay duration shows greater sensitivity to the availability of excitation, at least when estimated at relatively disinhibited regimes.

\begin{figure}[h!]
\begin{center}
\includegraphics[trim=0 0 0 0,width=15cm]{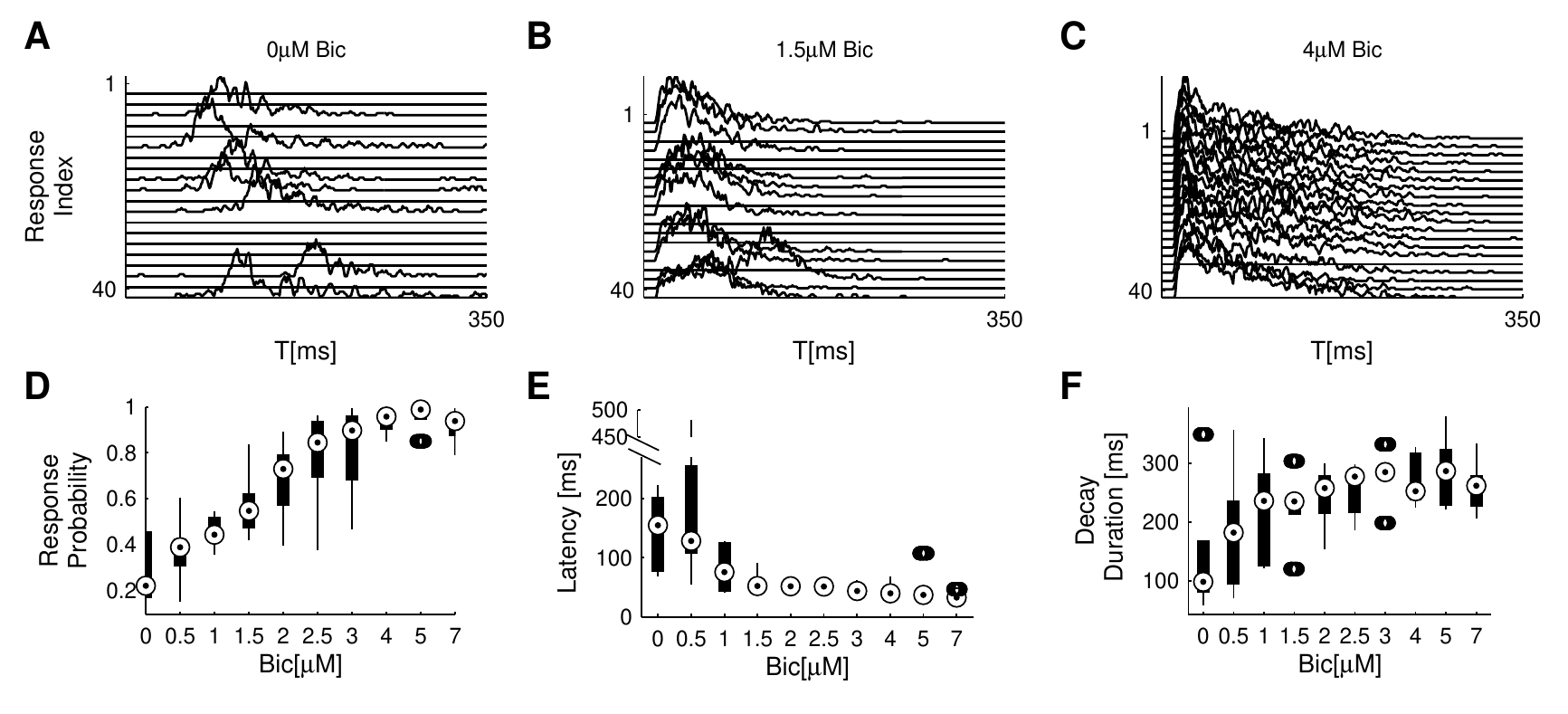}
\end{center}
\caption{\small{Effects of inhibitory transmission on network responsiveness: In panels {\bf A--C} consecutive network responses to repeating stimulation (1/6 sec$^{-1}$) at different Bicuculline doses from a single network are shown (to enhance visual clarity every other response was omitted). Panels {\bf D--F} summarize the sensitivities of response probability, response latency, and response decay duration to Bicuculine (six networks).  Box-plots represent the pooled distributions of local response probability ({\bf D}) and response latency ({\bf E}) collected from all six networks, each presented with 40 stimuli at ten increasingly applied doses. Out of  2400 responses, there are 40 outliers of local response probability, 80 outliers of latency and 200 outliers of decay duration.}}
\label{fg:DoseResBic}
\end{figure}

\begin{figure}[h!]
\begin{center}
\includegraphics[trim=0cm 0cm 0 0,width=15cm]{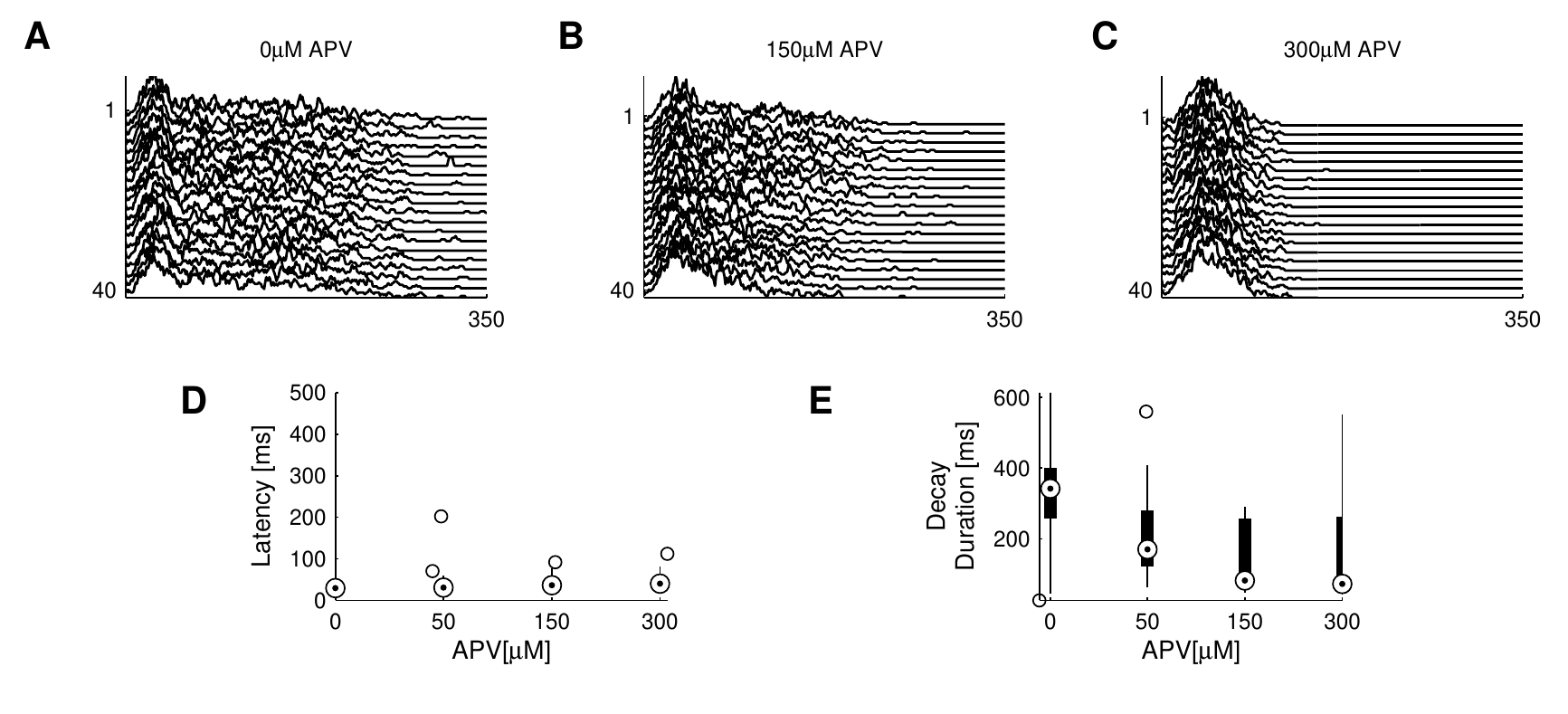}
\end{center}
\caption{\small{Effects of excitatory transmission on network responsiveness: APV is added on top of Bicuculine blockade (see Methods), inducing an extreme case of effectively excitatory networks. Panels {\bf A--C} show consecutive network responses to slow ($>1/7$ sec$^{-1}$) repeating stimulation at different APV doses from a single network; to enhance visual clarity every other response was omitted. Sensitivities of response latency and decay duration to APV are summarized in panels {\bf D--E} using box-plots representing the pooled distributions of latency ({\bf D}) and decay duration ({\bf E}) collected from four networks, each presented with 40 stimuli at four increasing doses. Out of  640 responses, there are 4 outliers of latency and a single excluded data point of decay duration.}}
\label{fg:DoseResAPV}
\end{figure}

Let us assume, based on the above results, that latency and decay duration represent aspects of inhibition and excitation. With this assumption in mind one might consider the use of the relations between these two measures of the individual response, as means to explore the stability (or, instability) of network excitation-inhibition ratio. The long term dynamics of both measures -- latency and decay -- as well as the interaction between them, may be intuited from the four examples shown in Figure 5.   These four examples (pulled out from a data set of 17 networks) represent the two extremes of a spectrum revealed by repeatedly stimulating networks at $1/3-1/5$ Hz over 50 minutes; this spectrum extends from narrow confined dynamics, through seemingly capricious, to well-organized several minutes-long trajectories across the latency-decay plane. Figure 5A shows the autocorrelation function of latency and decay series from two networks exhibiting oscillating autocorrelation of both measures, over minutes. Figure 5B represents latency-decay pairs obtained from individual responses of these two networks, forming well organized trajectories that persist many minutes. Although the autocorrelations of latency and decay in Figure 5A are highly coordinated, their ratio is not at all fixed, as may be inferred from the trajectories of Figure 5B. Rather, the ratio of latency to decay covers a wide dynamic range in their joint plane, markedly deviates from the main diagonal. (Note: the top panel of Figure 5A is calculated from the same network used to produce the population firing rate traces in Figure 1C.) The dynamics shown in Figure 5A and 5B represent a case that is extreme: these well-formed dynamics occupy relatively short segments of the entire data set. Most of the networks studied here did not exhibit coherent trajectories  that last as long. Rather, their dynamics ranged between long lasting correlations (hundreds of seconds, Figure 5C top inset) to no correlation at all (Figure 5C bottom inset).  Figure 5D demonstrates the latency-decay joint distribution from these networks and their vagrant responsiveness pattern; the colored lines depict segments of 50 consecutive responses, comparable with the colored trajectory-segments lengths presented in Figure 5B. The existence of long lasting correlations does not seem to be related to the spread of latency and decay duration of a given network. It might, however, be related to a higher center of mass along the decay axis. Overall, to the extent that latency and decay duration are proxies of inhibition and excitation, the data of Figure 5 implies that E/I should be treated as a dynamic variable over long temporal scales, rather than a parameter.

\begin{figure}[h!]
\begin{center}
\includegraphics[trim=0 0 0 0,width=15cm]{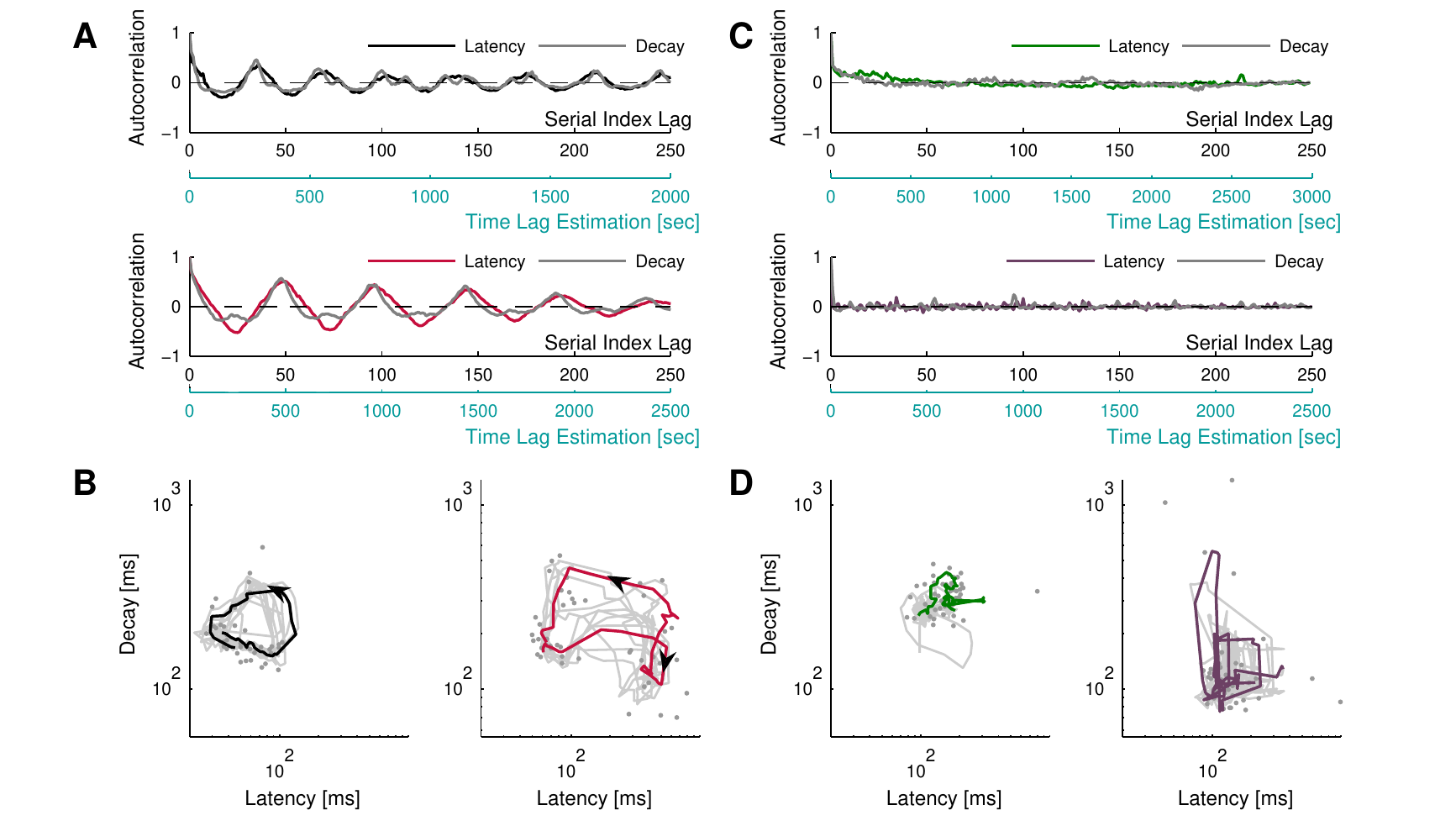}
\end{center}
\caption{\small{Latency-decay relations: Latency and decay duration were estimated from a data set of 17 networks during 50 minutes of repeating stimulation at 1/3--1/5 sec$^{-1}$. Examples for the different manners by which latency and decay duration behave and interact are presented. Panel {\bf A} depicts the autocorrelations of the time series of latency (black/red traces) and decay duration (gray traces) from two different networks, recorded over 50 minutes; both manifesting oscillations persisting over minutes. The autocorrelations are displayed with two lag-axes: the upper axis (black) displays gaps between serial numbers of consecutive successful responses (see Methods), while the lower axis (light blue) displays an estimation of the actual time lags, evaluated from the average inter-response-intervals. Pairs of latency and decay duration of individual responses, obtained from the two networks of panel {\bf A} (accordingly colored), are plotted using a log-log scale in panel {\bf B}, demonstrating organized activity of repeating trajectories on the latency-decay plane. The gray dots represent individual responses from a single trajectory, the colored curves depict the smoothed version of this trajectory (calculated using a moving average procedure), and the gray curves depict 50 minutes of the smoothed traces of latency-decay pairs from each network. Panel {\bf C} shows the autocorrelation of latency and decay duration time series from two other networks demonstrating the more common cases, where networks vagrantly explore the latency-decay plane with no apparent repeating trajectories. The latency series autocorrelations are depicted by the colored lines, and the decay series autocorrelations are depicted in gray. The top inset displays a case of long lasting correlations, while the bottom inset demonstrates a case of no correlation at all in both measures.  The latency-decay traces of these networks are displayed in panel {\bf D}, the colored lines demonstrate an apparently disorganized wandering across the latency-decay plane within a trace of 50 consecutive responses. }}
\label{fg:PhasePlaneNac}
\end{figure}

\section*{Discussion}

\noindent We provided experimental indications for the possible interpretation of network response variability in terms of a dynamical excitation-inhibition ratio. The sequence of our arguments begins with a description of correlations in time series of stimulus-evoked NSs occurrence, beyond the few seconds scale. We show that network responsiveness, as indicated by a local measure of response probability, correlates with the response latency. Aided by pharmacological manipulations, we point to links between the latency-decay ratio and excitation-inhibition ratio. We then reduce the multidimensional data of long-term population response variability to a ratio between two instantaneous, readily observable variables:  (1) the \textit{latency} from the stimulus time to the peak of the NS, and (2) the duration of \textit{decay} from the peak of the NS. Using these data, we demonstrate a spectrum of manners at which latency and decay duration behave and interact, indicating the dynamical nature of their ratio. 

Excitation-inhibition ratio is a determinant of network activity signature \cite{Brunel2000}. It is often assumed -- based on theoretical considerations -- that this ratio is balanced \cite{Shadlen1994,Vreeswijk1996,Shadlen1998,Vogels2005}. Experimental efforts to validate the balanced excitation-inhibition assumption are not conclusive: alongside supportive observations \cite{Haider2006,Okun2008,Xue2014}, the balanced E/I assumption was challenged by others \cite{Stevens1998,Wehr2003,Heiss2008}. We believe that the observations reported in the present study justify opening a discussion on the option of excitation-inhibition ratio being a dynamical variable that changes over minutes rather than a parameter that determines the nature of network activity in general, and its responsiveness in particular. Furthermore, we offer a method to instantaneously estimate the network state in terms of its E/I ratio directly at the population level. While not free of limitations, this estimation of E/I ratio is less sensitive to synaptic filtering introduced to estimations that are based on intracellular measurements. The latter are informative about the effective input at the single neuron level but remote from the neural population state.

We imagine at least three challenges that are entailed by the notion of excitation-inhibition ratio as a dynamical system variable, rather than a parameter. The \textit{first} challenge concerns modeling. The level of organization of the observed phenomenon (variability of population response) precludes models that focus on microscopic, this-or-that channel or synaptic receptor mechanisms underlying the rich network dynamics. In this respect it makes sense to use the terms `exciting forces' and `restoring forces' rather than `excitation' and `inhibition', thus shifting the load from local synaptic processes towards a richer repertoire of potential mechanisms that contribute to changes in network dynamics.  Models that are formulated in terms of global adaptation of both exciting and restoring forces and their interaction with time scales of stimulation and spontaneous activity are wanted.  Such models might shed light on possible connections between the dynamics of stimulus evoked responses reported here and the well-documented complex statistics of spontaneous series of network spikes \cite{Segev2002,Beggs2003,Wagenaar2006c,Mazzoni2007}.  The \textit{second} challenge we envision is related to functional aspects of network activity in general, and the impacts of slow dynamics of exciting and restoring forces on the efficacy of different representation schemes, in particular.  Whether such schemes involve population or spike-time `neural codes', their sensitivity to slow network dynamics should be considered.  One might imagine scenarios where slow network dynamics of the kind described here can have constructive impacts on the ability of neural systems to explore and adapt to a changing environment. The \textit{third} challenge is even more general; it concerns the inherent tension between approaches that heavily rely on structural measures and approaches that are more concerned with `effective' measures in providing insightful information on neural systems dynamics and function. The multiple timescales of network responsiveness as well as the dynamic exciting-restoring forces emerge from an allegedly stable structure, at least in terms of the number of excitatory and inhibitory neurons and the number of their synapses.  Indeed, it is difficult to imagine the dynamics over minutes as presented here being determined by ongoing changes in structural network parameters; in this respect our results further contribute to recent calls that challeng attempts to relate the structure of neural networks to their dynamics \cite{Marder2014}.    

Whether or not our interpretations that are based on \textit{in-vitro} experimental analyses may be generalized, the very possibility of network-level exciting and restoring forces being dynamical variables seems to deserve serious consideration by those interested in theoretical and applied aspects of neural response variation.

\bibliography{NH_SM_2014}
\end{document}